\documentclass[showkeys,pra,tightenlines,12pt]{revtex4-1}

\usepackage{epsfig}
\usepackage{graphicx}
\usepackage{amsmath}

\usepackage{amsthm}
\usepackage{amsfonts}
\usepackage{subfigure}
\usepackage[english, ruled]{algorithm2e}

\newtheorem{Definition}{Definition}
\newtheorem{Lemma}{Lemma}
\newtheorem{propsIso}{Property}
\newtheorem{propsTriang}{Property}

\newtheorem*{Remark}{Remark}

\def\Dist#1#2{\mathcal{D} \left( #1, #2 \right) }
\def\Trace{\textrm{Tr}}

\def\ket#1{|#1 \rangle}

\def\ketbra#1#2{|#1 \rangle \langle #2|}
\def\bradensket#1#2#3{\langle #1| #2 |#3 \rangle}
\def\dens#1{\ketbra{#1}{#1}}

\def\NQ{$n$-qubits }
\def\cos{\mathrm{cos}}
\def\sin{\mathrm{sin}}

\def\IsoInd#1#2{\textit{Iso}_#1\left(#2\right)}

\begin{document}

\title{Characterizing error propagation in quantum circuits: the Isotropic Index.}

\author{Andr\'e L. Fonseca de Oliveira}
\email{fonseca@ort.edu.uy}
\affiliation{Facultad de Ingenier\'ia, Universidad ORT Uruguay, Uruguay.}

\author{Efrain Buksman}
\email{buksman@ort.edu.uy}
\affiliation{Facultad de Ingenier\'ia, Universidad ORT Uruguay, Uruguay.}

\author{Ilan Cohn}
\email{icohn@uni.ort.edu.uy}
\affiliation{Facultad de Ingenier\'ia, Universidad ORT Uruguay, Uruguay.}

\author{Jes\'us Garc\'ia L\'opez de Lacalle}
\email{jglopez@etsisi.upm.es}
\affiliation{ETSISI, Universidad Polit\'ecnica de Madrid, Spain.}

\begin{abstract}
This paper presents a novel index in order to characterize error propagation in quantum circuits by separating the resultant mixed error state in two components: an isotropic component, that quantifies the lack of information, and a dis-alignment component, that represents the shift between the current state and the original pure quantum state. The Isotropic Triangle, a graphical representation that fits naturally with the proposed index, is also introduced. Finally, some examples with the analysis of well-known quantum algorithms degradation are given.

\end{abstract}

\maketitle

\section{Introduction}
\label{sec_Intro}

The design and construction of reliable quantum computers is one of the greatest  
challenges of these decades. Long-lived quantum state superposition, entanglement maintenance and manipulation of large-qubits systems require the analysis of several technological implementation aspects of quantum systems \cite{Ladd_2010}.

One of the most important problems is that quantum systems cannot be completely isolated from the environment. This fact, together with imperfections in gate applications, state preparations and others, induces errors in any quantum computation \cite{Knill_1998}\cite{Gottesman_2010}\cite{Devitt_2013}. While there is a fault-tolerant model of quantum computing based on the correction of errors below a certain threshold \cite{Gottesman_1998}\cite{Aharonov_2008} this method is very expensive in computational resources. Therefore, the development of new tools in order to analyze the effect and propagation of quantum errors is an important issue.

This paper presents a novel index, the Isotropic Index, useful to identify uncorrectable components of mixed quantum states, the isotropic component; and the dis-alignment with respect to an original reference pure state.

The structure of the paper is as follows. Isotropic errors model and the concept of isotropic error state are introduced in section \ref{sec_IsoErr}. In section \ref{sec_IsoIndex} we define the Isotropic Index and its graphical representation: the Isotropic Triangle. This representation gives a new perspective in the analysis of the performance loss in quantum algorithms. The relation between the isotropic states defined in \cite{Horodecki_1999} and isotropic error states is presented in section \ref{sec_EstIsoHor}. 

In later sections, \ref{sec_Grover} and \ref{sec_Shor}, the index is used to explain the degradation of results in two well-known quantum algorithms: Grover's quantum search \cite{Grover_1997} and Shor's quantum error-correcting  code \cite{Shor_1995}. In both cases depolarizing channel models are used.

\section{Isotropic errors}
\label{sec_IsoErr}

In order to define an isotropic error state, a quantum state obtained from a time-independent isotropic error process, it is useful to introduce the definition of such error model as in \cite{GarciaLopez_2003A} and \cite{GarciaLopez_2007A}.

\begin{Definition}[\bf{Time-independent isotropic error model}]
\label{def_IsoErr}
A time-independent isotropic error process, with an \NQ pure quantum state $\ket{\psi}$ as the reference state, is a stochastic process $f(\ket{\psi} , 0, t)$, $t \ge 0$, that holds:
\begin{itemize}
\item At $t = 0$, the density probability distribution is $f(\ket{\psi} , 0, 0) = \delta (\ket{\psi})$, i.e. there is no error and the state is $\ket{\psi}$ with probability equal to $1$.

\item The density probability distribution $f(\ket{\psi} , 0, t_k)$ is isotropic for all $t_k$. Let $\ket{\psi(t_k)}$ be the random variable at $t_k$. The probability of obtaining the state $\ket{\phi}$ depends only on the quantum distance between $\ket{\psi}$ and $\ket{\phi}$,
\begin{equation}
\label{eqn_IsoErr_ISoProb}
P \left( \ket{\psi(t_k)} = \ket{\phi} \right) = F ( \Dist{\ket{\psi}}{\ket{\phi}} ).
\end{equation}

\item For all $t_k$ and $t_{k+1}$, $t_{k} < t_{k+1}$, $f(\ket{\psi}, 0, t_{k})$ and $f( \ket{\psi}, t_{k}, t_{k+1} )$ are independent.
\end{itemize} 
  
\end{Definition}

With the purpose of obtaining a closed form for an isotropic error state, we consider the
reference state (r.s.) as the basis state $\ket{0}$ (for a general formulation with r.s. $\ket{\psi}$, one only needs a basis change).

According to the definition a ensemble of states
\begin{equation}
\label{eqn_IsoErr_ErrIsoVec}
\ket{\psi(t_k)} = a_0(k) \ket{0} + a_1(k) e^{i \varphi_1(k)} \ket{1} + \dots + a_{2^n-1}(k) e^{i \varphi_{2^n-1}(k)} \ket{2^n-1},
\end{equation}
is an (mixed) isotropic error state (r.s. $\ket{0}$) if holds
\begin{itemize}
\item $\ket{i}$, $0 \le i \le 2^n-1$, the canonical basis vectors (decimal notation),
\item $\varphi_i(k)$ random variables with uniform distribution in $[0, 2\pi)$ (the distance does not depend on the relative phases) and
\item $a_i(k)$, random variables in $[0, 1]$ subject to the constraints
\begin{itemize}
\item $\forall k$, $\sum_{i=0}^{2^n-1} a_i^2(k)= 1$,
\item $a_i(k)$, $1 \leq i \leq 2^n - 1$, have the same probability distribution.
\end{itemize}
\end{itemize}

\begin{Remark}
In this case both the distance $\Dist{\ket{0}}{\ket{\psi(t_k)}}$ and the probability distribution depend only on $a_0(k)$.
\end{Remark}

\subsection{Density operators}
\label{subsec_IsoErr_Dens}

The density matrix corresponding to the quantum state in (\ref{eqn_IsoErr_ErrIsoVec}) is
\begin{equation}
\label{eqn_IsoErr_MatDens}
\rho = E \left[ \, \dens{\psi(t_k)} \, \right] = E \left[ \left( \gamma_{ij}(k) \right) \right],
\end{equation}
where $E[\cdot]$ is the expected value of the random matrix and, for $1 \leq i, j \leq 2^n -1$,
\begin{itemize}
\item $\gamma_{ll}(k) = a_l^2(k)$, $\forall i = j = l $,
\item $\gamma_{0m}(k) = a_0(k) a_m(k) e^{-i \varphi_m(k)}$, $\forall j = m \neq 0$,
\item $\gamma_{l0}(k) = a_l(k) a_0(k) e^{i \varphi_l(k)}$, $\forall i = l \neq 0$,
\item y $\gamma_{lm}(k) = a_l(k) a_m(k) e^{i \left( {\varphi_l(k) - \varphi_m(k)} \right)}$, $\forall l \neq m$.
\end{itemize}

As a direct consequence of the hypothesis made in (\ref{eqn_IsoErr_ErrIsoVec}), one has 
\begin{itemize}
\item $E\left( e^{i \varphi_i(k)} \right) = 0$,
\item $E\left( a_i(k) a_j(k) e^{i \left( \varphi_i(k)-\varphi_j(k) \right)} \right) = E\left( a_i(k) a_j(k) \right) E\left( e^{i \varphi_i(k)} \right) E\left( e^{-i \varphi_j(k)} \right) = 0$
\item $E\left( a_0(k) \right) = \lambda_0$,
\item $E\left( a_i(k) \right) = \lambda_1, \forall \, i, \, 1 \leq i \leq 2^n -1$,
\item $(2^n-1) \lambda_1 + \lambda_0 = 1$.
\end{itemize}

Therefore, the mixed density matrix $\rho_{iso}$ representing an isotropic error state (with r.s. $\ket{0}$) has the form
\begin{equation}
\label{eqn_IsoErr_DensIso}
\rho_{iso} = 
\left[ \begin{array}{cccc}
  \lambda_0 & 0 & 0 & 0 \\
  0 & \lambda_1 & 0 & 0 \\
  \vdots & \vdots & \ddots & \vdots \\
	0 & 0 & \cdots & \lambda_1
\end{array} \right].
\end{equation}

\begin{Definition}[\bf{Isotropic error state}]
\label{def_IsoErr_Dens}
Let $\ket{\psi}$ be a reference state and $M$ a basis change matrix so that $\ket{0} = M \ket{\psi}$. A mixed state $\rho$ is an isotropic error state (r.s.$\ket{\psi}$) if the the state $\bar{\rho} = M \rho M^\dagger$ has the form (\ref{eqn_IsoErr_DensIso}) and can be expressed as
\begin{equation}
\bar{\rho} = \left\{ \begin{array}{lcr}
\left( 2^n \lambda_1 \right) \frac{I}{2^n}  + \left( \lambda_0 - \lambda_1 \right) \rho_0, & & \lambda_0 \ge \lambda_1 \\ \\
\left( 2^n \lambda_0 \right) \frac{I}{2^n} + \left( 2^n - 1\right) \left( \lambda_1 - \lambda_0 \right) \rho_{N0}, & & \lambda_0 < \lambda_1
\end{array} \right. ,
\label{eqn_IsoErr_DensIsoDecomp}
\end{equation}
where $\rho_0 = \dens{0}$ and
\begin{equation}
\label{eqn_IsoErr_MezclaIsoOrt}
\rho_{N0} = \frac{\left(I - \rho_0 \right)}{\left( 2^n - 1\right)} = \frac{1}{\left( 2^n - 1\right)} \left[ \begin{array}{cccc}
  0 & 0 & 0 & 0 \\
  0 & 1 & 0 & 0 \\
  \vdots & \vdots & \ddots & \vdots \\
	0 & 0 & \cdots & 1
\end{array} \right]
\end{equation}
the orthogonal isotropic mixed state relative to the reference state.
\end{Definition}

\section{Isotropic Index}
\label{sec_IsoIndex}

In \cite{Fonseca_2011} a double index that characterizes error isotropy in one-qubit states was proposed. This index has some restrictions: firstly it has a geometric, but not quantum, interpretation. Two different ensembles of states represented by the same density operator are indistinguishable by quantum measurements. Second, although it can be generalized for \NQ states, it has a hard, and ill conditioned, computation.

In this section we define a new double index that quantifies how far is an arbitrary \NQ mixed state from being an isotropic error state: the Isotropic Index. We also introduce a special representation, a triangular graph, for this index. 

\subsection{Mixed state decomposition}
\label{subsec_IsoIndex_Descomp}

In order to characterize the isotropic component of an arbitrary mixed state, it is necessary to decompose the state into a similar form as in (\ref{eqn_IsoErr_DensIsoDecomp}).

\begin{Lemma}
For any given \NQ quantum state $\rho$ the following decomposition is always possible
\begin{equation}
\label{eqn_IsoErr_DensDecomp}
\rho = p \frac{I}{2^n} + \left( 1 -p \right) \hat{\rho},
\end{equation}
where $p$ is a probability, $I$ is the identity matrix and $\hat{\rho}$ is a density matrix with at least one null eigenvalue. If $\rho$ already has one null eigenvalue, then $p = 0$.

\begin{proof}
Since $\rho$ is a density matrix, there is always a unitary matrix $M$ that diagonalizes $\rho$ into $\rho_d$,
\begin{equation}
\rho_d = M^\dagger \rho M = \lambda I + \Theta = 2^n \lambda \frac{I}{2^n} + \Trace \left( \Theta \right) \hat{\rho_d} = p \frac{I}{2^n} + \left(1 - p \right) \hat{\rho_d},
\end{equation} 
with $\lambda$ being the smaller eigenvalue of $\rho$. Then
\begin{equation}
\rho = M \rho_d M^\dagger = p \frac{I}{2^n} + \left(1 - p \right) M \hat{\rho_d} M^\dagger =p \frac{I}{2^n} + \left(1 - p \right) \hat{\rho} .
\end{equation} 
\end{proof}
\end{Lemma}

\subsection{Isotropic Index definition}
\label{subsec_IsoIndex_Index}

Considering the definition \ref{def_IsoErr_Dens} and the former decomposition, it is easy to see that equation (\ref{eqn_IsoErr_DensIsoDecomp}) is a particular case of (\ref{eqn_IsoErr_DensDecomp}). Therefore, a quantum state $\rho$ is an isotropic error state (r.s. $\ket{0}$) if when decomposing the state as in (\ref{eqn_IsoErr_DensDecomp}) the resulting state $\hat{\rho}$ is either $\rho_0$ or $\rho_{N0}$. For any other reference state a change of basis needs to be performed.

\begin{Definition}[\bf{Isotropic Index}]
\label{def_IsoErr_DefIndIso}
Considering the pure quantum state $\rho_{\varphi} = \dens{\varphi}$ as the reference, the Isotropic Index for an arbitrary state $\rho$ is defined as the double index
\begin{equation} 
\label{eqn_IsoErr_DefIndIso}
\IsoInd{\varphi}{\rho} = \left( A, p \right)
\end{equation}
being
\begin{itemize}
\item $A$, the \emph{Isotropic Alignment}, defined as
\begin{equation} 
\label{eqn_IsoErr_DefIndIsoAlin}
A = Fid \left( \hat{\rho}, \rho_{\varphi} \right) - Fid \left( \hat{\rho}, \rho_{N\varphi} \right)
\end{equation}
where $Fid$ is the fidelity between quantum states \cite{Nielsen_2000A}, $\hat{\rho}$ comes from the decomposition (\ref{eqn_IsoErr_DensDecomp}) and $\rho_{N\varphi} = \left(I - \dens{\varphi} \right) / (2^n-1)$ (orthogonal isotropic mixed state of $\rho_{\varphi}$),
\item and $p = 2^n \lambda$, the \emph{Isotropic Weight}, with $\lambda$ being the smaller eigenvalue of $\rho$.
\end{itemize}

\end{Definition}

\subsubsection{Isotropic Index properties}
\label{subsubsec_IsoIndex_IndexProps}

\begin{propsIso}
\label{prop_IsoIndex_IndexProps1}
Since $p$ is defined as a probability it is always in $[0,1]$. In the case where $p=1$, the state $\rho$ is equal to $I/2^n$, $\hat{\rho}$ has no meaning and the \emph{Isotropic Alignment} $A$ has a default value of $1$. In this case, $\rho$ is an isotropic error state.
\end{propsIso}

\begin{propsIso}
\label{prop_IsoIndex_IndexProps2}
The \emph{Isotropic Alignment} $A$ lies in $[-1,1]$, since the states $\rho_{\varphi}$ and $\rho_{N\varphi}$ are orthogonal to each other. For all \emph{Isotropic Weight} $p$, $\rho$ is an isotropic error state if either $A = 1$ or $A = -1$.
\end{propsIso}

\begin{propsIso}
\label{prop_IsoIndex_IndexProps3}
The index is invariant under any unitary operator $U$ applied to both the studied state and the reference state, i.e. $\IsoInd{\varphi}{\rho} = \IsoInd{\psi}{\sigma}$ with $\ket{\psi} = U \ket{\varphi}$ and $\sigma = U \rho U^\dagger$.
\end{propsIso}

\begin{propsIso}
\label{prop_IsoIndex_IndexProps4}
Given any trace-preserving quantum operation, represented via its Kraus' operators as
\begin{equation} 
\label{eqn_IsoErr_QOpKraus}
\varepsilon(\rho) = \sum_{i=1}^k M_i \rho M_i^\dagger, \quad \textrm{where} \quad \sum_{i=1}^k M_i^\dagger M_i = I,
\end{equation}
the \emph{Isotropic Weight} $p$ is non-decreasing if $\varepsilon(\rho)$ is \emph{unital} \cite{Bourdon_2004}
\begin{equation} 
\label{eqn_IsoErr_CondAnchoIsoKraus}
\sum_{i=1}^k M_i M_i^\dagger = I.
\end{equation}

\begin{proof}
Let $\rho$ be the quantum state studied with its corresponding decomposition as in equation(\ref{eqn_IsoErr_DensDecomp})
\begin{displaymath}
\rho = p \frac{I}{2^n} + (1-p) \hat{\rho}.
\end{displaymath}

Then
\begin{equation} 
\label{eqn_IsoErr_CondAnchoIsoKrausOp}
\varepsilon(\rho) = \frac{p}{2^n} \sum_{i=1}^k M_i M_i^\dagger + (1-p) \sum_{i=1}^k M_i \hat{\rho} M_i^\dagger = p \frac{I}{2^n} + (1-p) \sum_{i=1}^k M_i \hat{\rho} M_i^\dagger.
\end{equation}

Hence, the \emph{Isotropic Weight} is $p + (1-p)\alpha$, where $\alpha \geq 0$ is the \emph{Isotropic Weight} of $\sum_{i=1}^k M_i \hat{\rho} M_i^\dagger$. As can be observed from the former expression, the condition given by (\ref{eqn_IsoErr_CondAnchoIsoKraus}) is sufficient, yet not necessary.
\end{proof}
\begin{Remark}
For closed quantum operations (described by a unitary operator) the \emph{Isotropic Weight} remains constant.
\end{Remark}
\end{propsIso}

\subsection{Graphical representation: the Isotropic Triangle}
\label{subsec_IsoIndex_IsoTriang}

As stated in property \ref{prop_IsoIndex_IndexProps1}, when $p=1$ the index $\IsoInd{\varphi}{\rho} = \left( A, 1 \right)$ represents the same point in space for all $A$. For this reason the most appropriate form of representation is a triangle.

Let the Isotropic Index be $\IsoInd{\varphi}{\rho} = (A,p)$. The coordinates on the triangle are determined by the map
\begin{eqnarray}
x & = & (1-p) A, \label{eqn_IsoIndex_IsoTriang_CoordX} \\
y & = & p. \label{eqn_IsoIndex_IsoTriang_CoordY}
\end{eqnarray}

\begin{figure}[!htb]
\centering
\includegraphics[width=0.6\textwidth,keepaspectratio=true]{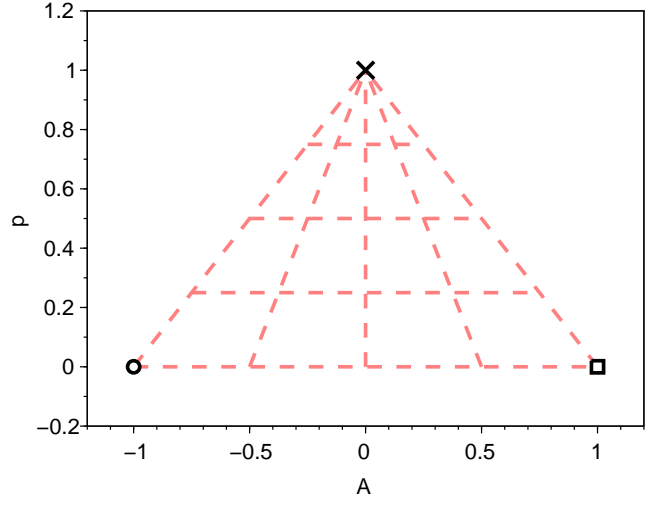}
\caption{Isotropic Triangle. This figure shows the double Isotropic Index for the states $\rho_{\varphi}$ (square), $\rho_{N\varphi}$ (circle) and $I/2^n$ (cross).}
\label{fig_IsoIndex_IsoTriang_Def}
\end{figure}

Figure \ref{fig_IsoIndex_IsoTriang_Def} shows the reference state ($\rho_{\varphi}$), the orthogonal isotropic mixed state ($\rho_{N\varphi}$) and the maximally mixed state $I/2^n$ in the Isotropic Triangle.

\subsubsection{Isotropic Triangle properties}
\label{subsubsec_IsoIndex_TriangProps}

Given the invariance of the Isotropic Index with respect to a change of basis, on the following properties we consider the r.s. as $\ket{0}$.

\begin{propsTriang}[\bf Null probability states]
\label{prop_IsoIndex_TriangProps1}
Considering the projector constructed by the r.s., states with null probability ($Fid(\rho_0, \rho)=0$), lie at the bottom of the triangle in the interval $\left[-1, -\sqrt{\frac{1}{2^n-1}}\right]$.
\begin{proof}

With the decomposition (\ref{eqn_IsoErr_DensDecomp}), the respective probability is at least $p/(2^n)$. This implies that for null probability the \emph{Isotropic Weight} must be $p = 0$ and, therefore, $\rho = \hat{\rho}$. Since $Fid \left( \hat{\rho}, \rho_0 \right) = 0$ (square root of the probability), in this case we have that $A = - Fid \left( \hat{\rho}, \rho_{N0} \right)$. If $\rho = \rho_{N0} \Rightarrow A = -1$. Hence, the states with minimum fidelity with respect to $\rho_{N0}$ and null probability are the $2^n - 1$ canonical basis states (without $\ket{0}$). Thus, 
\begin{equation}
\hat{\rho} = \rho_i = \ketbra{i}{i} , \ Fid \left( \hat{\rho}, \rho_{N0} \right) = \sqrt{\bradensket{i}{\rho_{N0}}{i}} = \frac{1}{\sqrt{2^n-1}}.
\label{eqn_IsoIndex_TriangProps1_A1}
\end{equation}
\end{proof}
\end{propsTriang}

\begin{propsTriang}[\bf Pure states isotropy]
\label{prop_IsoIndex_TriangProps2}
Let $\rho$ be the density matrix of a pure state of the form $\ket{\psi} = a_0 \ket{0} + a_1 e^{i \varphi_1} \ket{1} + \dots + a_{2^n-1} e^{i \varphi_{2^n-1}} \ket{2^n-1}$. In this case, it is immediate to see that $p=0$, since there is a unique eigenvalue equal to $1$. Hence, the calculation of $A$ reduces to
\begin{equation}
Fid_0 = Fid \left( \ket{\psi}, \ket{0} \right) = a_0,
\label{eqn_IsoIndex_TriangProps2_A1}
\end{equation}

\begin{eqnarray}
Fid_{N0} & = & Fid \left( \ket{\psi}, \rho_{N0} \right) = \sqrt{\bradensket{\psi}{\rho_{N0}}{\psi}}, \nonumber \\
& = & \sqrt{\sum_{k=1}^{2^n-1} \frac{a_k^2}{2^n-1}} = \sqrt{\frac{1 - a_0^2}{2^n-1}},
\label{eqn_IsoIndex_TriangProps2_A2}
\end{eqnarray}

\begin{equation}
\Rightarrow A = Fid_0 - Fid_{N0} = a_0 - \sqrt{\frac{1 - a_0^2}{2^n-1}}.
\label{eqn_IsoIndex_TriangProps2_A3}
\end{equation}

Therefore, pure states are located in the bottom of the triangle in the interval $\left[-\sqrt{\frac{1}{2^n-1}}, 1 \right]$.
\end{propsTriang}

\begin{Remark}
Since the calculated probability depends only on $a_0^2$, all pure states with the same probability have the same index. 
\end{Remark}

\begin{equation}
\IsoInd{0}{\rho} = \left(a_0 - \sqrt{\frac{1 - a_0^2}{2^n-1}} , 0 \right).
\label{eqn_IsoIndex_TriangProps2_A4}
\end{equation}

\begin{figure}[htpb]
\centering
\includegraphics[width=0.6\textwidth,keepaspectratio=true]{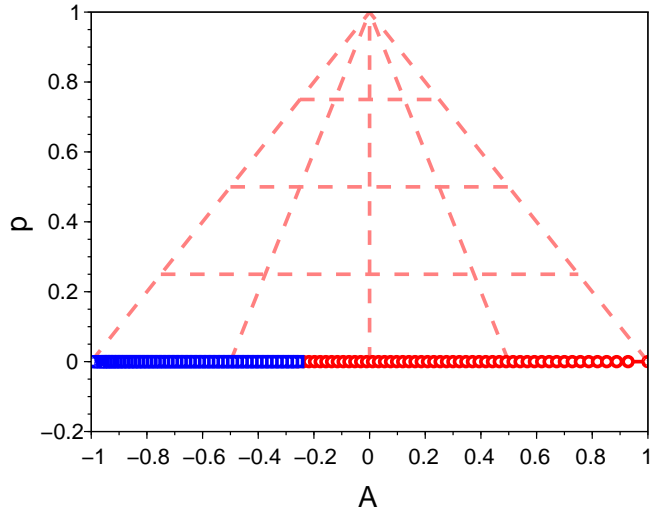}
\caption{Properties 1 and 2. Pure sates (red circles) and null probability states (blue squares).}
\label{fig_IsoIndex_TriangProps1y2} 
\end{figure}

Figure \ref{fig_IsoIndex_TriangProps1y2} illustrates the pure states (black) and the null probability (gray) zones. Note that pure states with null probability have the index in the intersection of both zones, $\IsoInd{0}{\rho} = \left(-\frac{1}{\sqrt{2^n-1}},0 \right)$.

\begin{figure}[htpb]
	\centering
	\subfigure[Index evolution due to a depolarizing channel (error probabilities from $0$ to $1$). Initial states: reference state (black circles), null probability state (blue squares) and random state with non-null \emph{Isotropic Weight} (red crosses).]{
	\label{fig_IsoIndex_TriangProps3} 
	\includegraphics[width=0.45\textwidth,keepaspectratio=true]
	{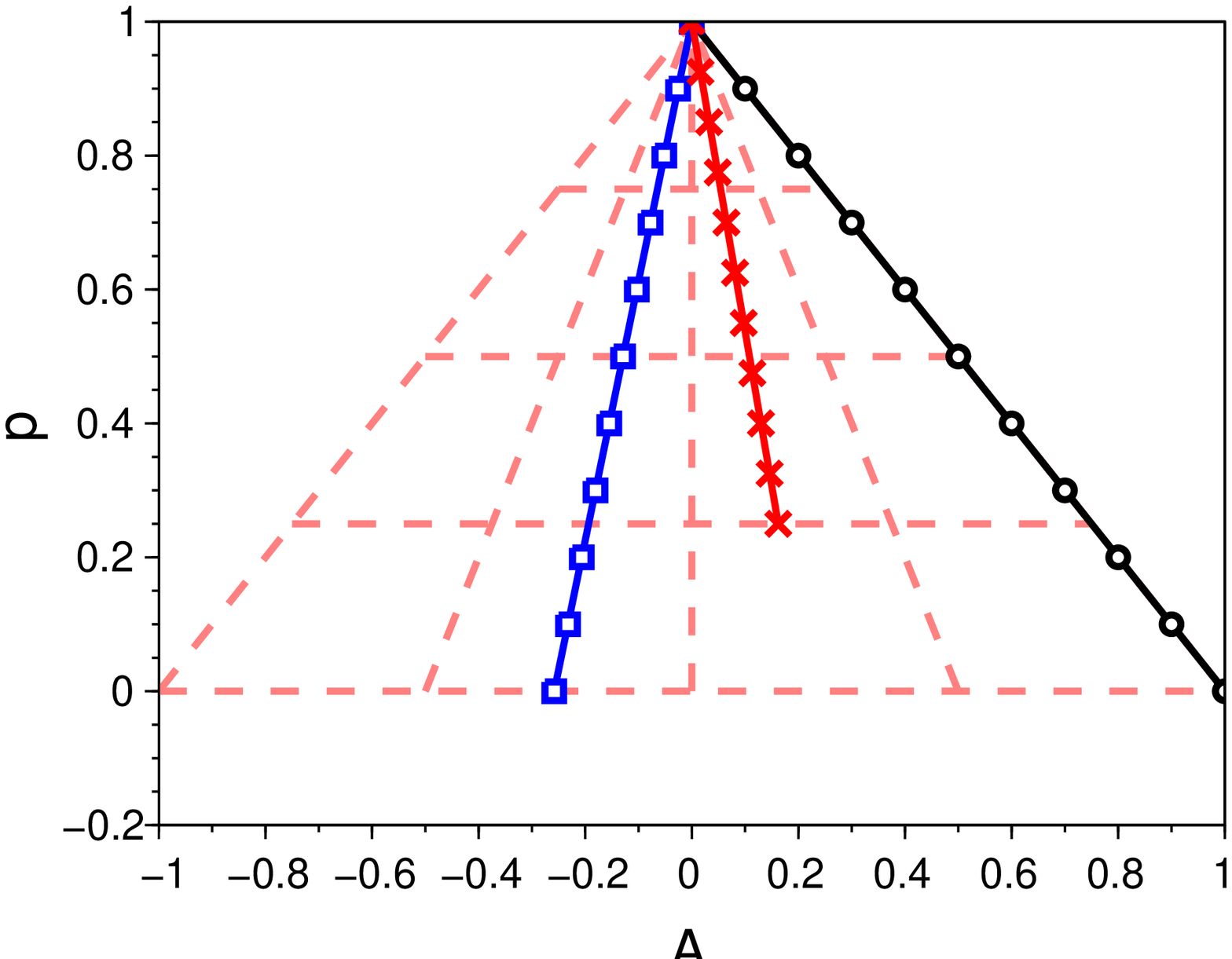}}
	\hfill
	\subfigure[Index evolution due to random unitary evolution. Initial states: reference state (black circles), random state with $p = 0.25$ (blue squares) and random state with $p = 0.5$ (red crosses).]{
	\label{fig_IsoIndex_TriangProps4} 
	\includegraphics[width=0.45\textwidth,keepaspectratio=true]
	{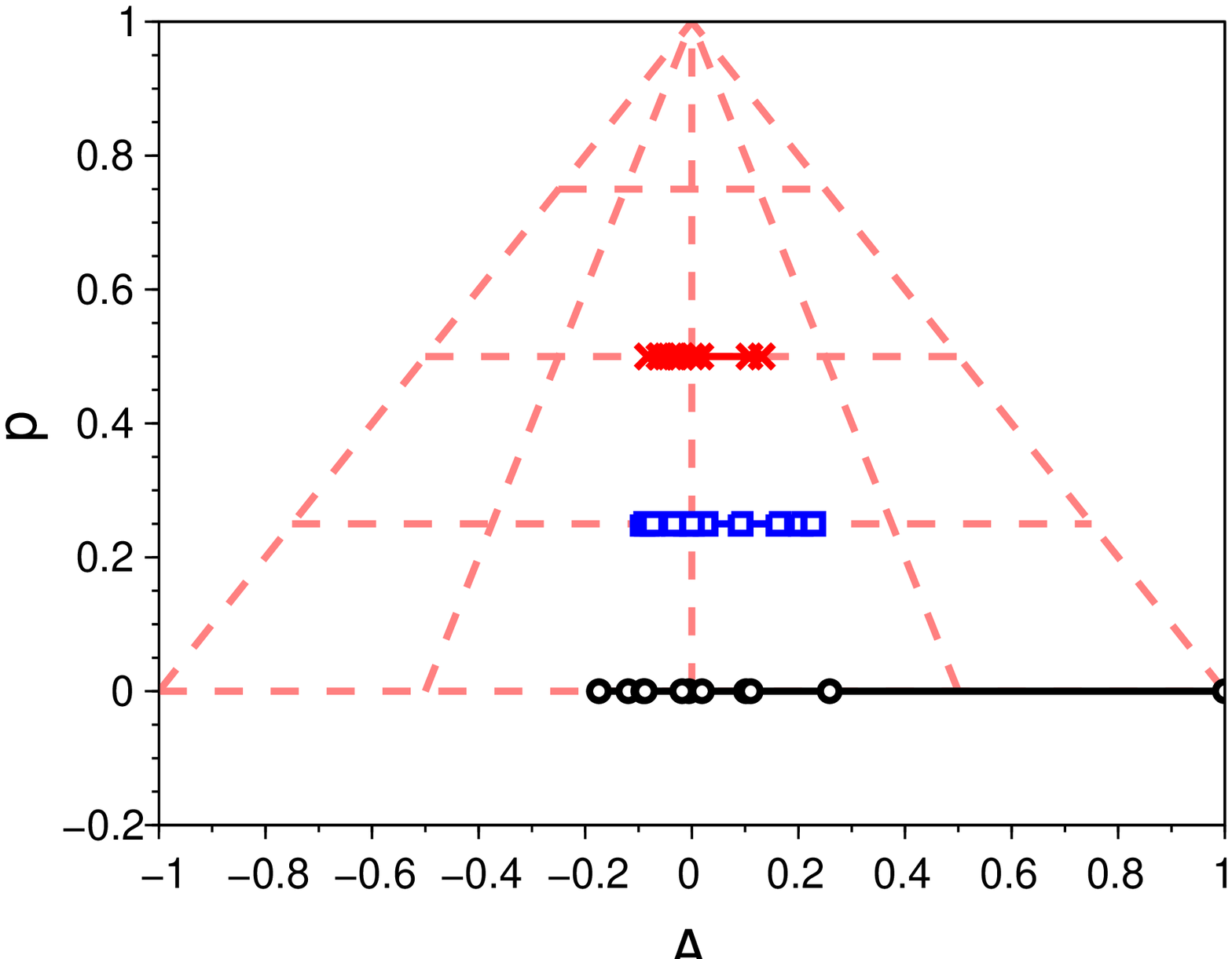}}
	\caption{Examples for properties \ref{prop_IsoIndex_TriangProps3} and \ref{prop_IsoIndex_TriangProps4}, for $4$-qubits states.}
	\label{fig_IsoIndex_TriangProps3y4} 
\end{figure}

\begin{propsTriang}[\bf Depolarizing channel]
\label{prop_IsoIndex_TriangProps3}
Let $\rho$ be an initial state with a decomposition as in (\ref{eqn_IsoErr_DensDecomp}) and an index given by $\IsoInd{0}{\rho} = \left(A,p\right)$. After applying a depolarizing channel operator error \cite{Nielsen_2000A} the resulting state is
\begin{equation}
\varepsilon \left( \rho \right) = \gamma \frac{I}{2^n} + \left( 1 - \gamma \right) \rho, 
\label{eqn_IsoIndex_TriangProps3_A1}
\end{equation}
with $\gamma \in [0, 1]$. In this case 
\begin{eqnarray}
\varepsilon \left( \rho \right) & = & \gamma \frac{I}{2^n} + \left( 1 - \gamma \right) p \frac{I}{2^n} + \left( 1 - \gamma \right) \left( 1 - p \right) \hat{\rho} \nonumber \\
& = & \left( \gamma + p - \gamma p \right) \frac{I}{2^n} + \left( 1 - \gamma - p + \gamma p \right) \hat{\rho} \nonumber \\
& = & \alpha \frac{I}{2^n} + \left( 1 - \alpha \right) \hat{\rho},
\label{eqn_IsoIndex_TriangProps3_A2}
\end{eqnarray}
being $\alpha$ the new \emph{Isotropic Weight}. Then $A$ remains constant (same $\hat{\rho}$) and the \emph{Isotropic Weight} varies from $p$ ($\gamma = 0$) to $1$ ($\gamma = 1$). As can be seen in figure \ref{fig_IsoIndex_TriangProps3}, the trajectories are line segments joining the initial state with the maximally mixed state ($\rho = \frac{I}{2^n}$).
\end{propsTriang}

\begin{propsTriang}[\bf Unitary evolution]
\label{prop_IsoIndex_TriangProps4}

Due to property \ref{prop_IsoIndex_IndexProps4}, for a fixed reference state, if the studied state evolves within a closed system (unitary operator model), the \emph{Isotropic Weight} remains constant. Figure \ref{fig_IsoIndex_TriangProps4} shows the evolution of the index for some initial states. 
\end{propsTriang}

\begin{figure}[ht!]
\centering
\subfigure[Constant probability zones: $0.8$ (black), $0.6$ (dark gray), $1/2^n$ (medium gray) and $0.25/2^n$ (light gray).]{
\label{Fig_IndIso_PROB_VarProbs_Pfijas_4qb}
\includegraphics[width=0.45\textwidth]{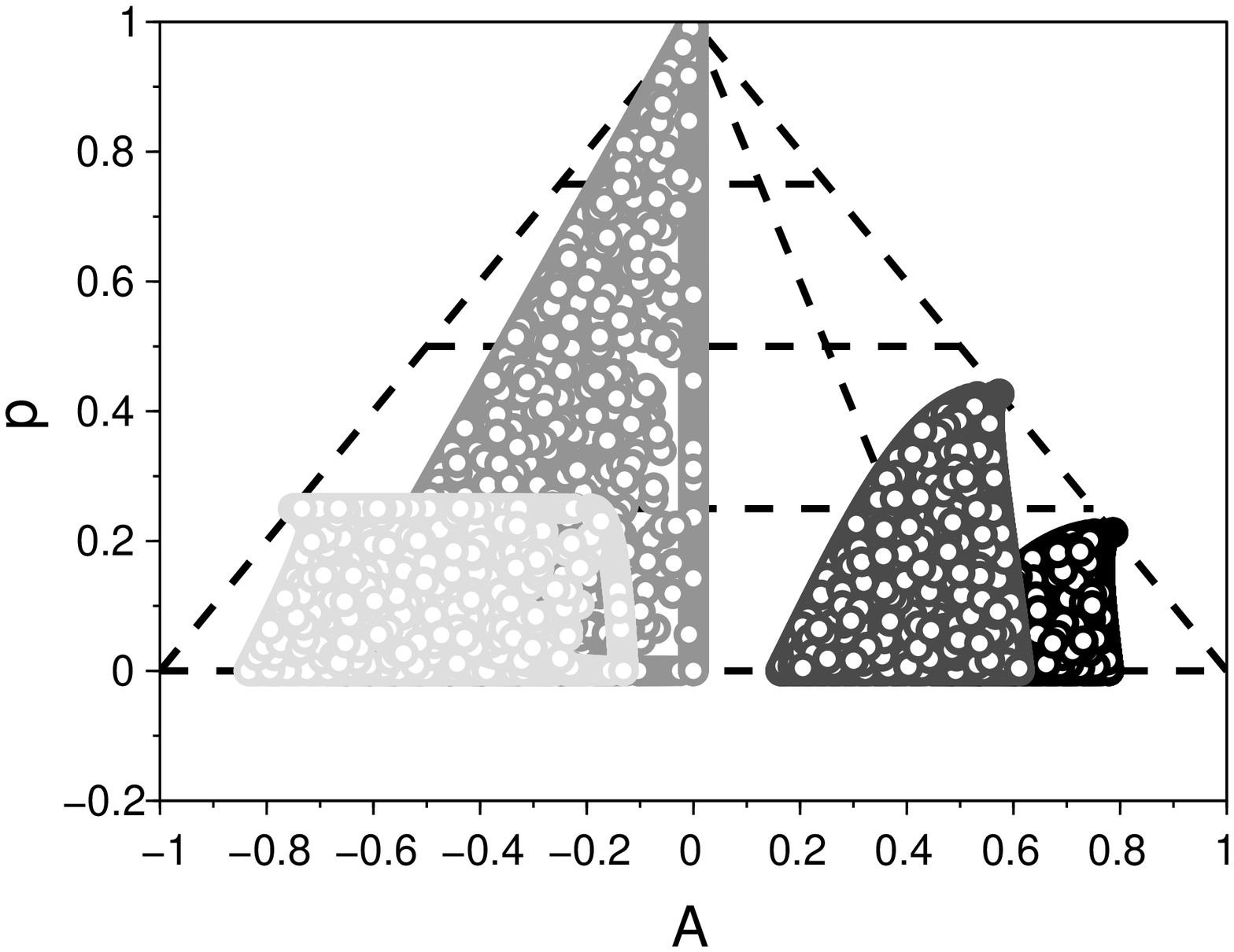}}
\hfill
\subfigure[Random states probabilities. Darker zones have higher probabilities.]{
\label{Fig_IndIso_PROB_VarProbs_EstRand_4qb} 
\includegraphics[width=0.45\textwidth]{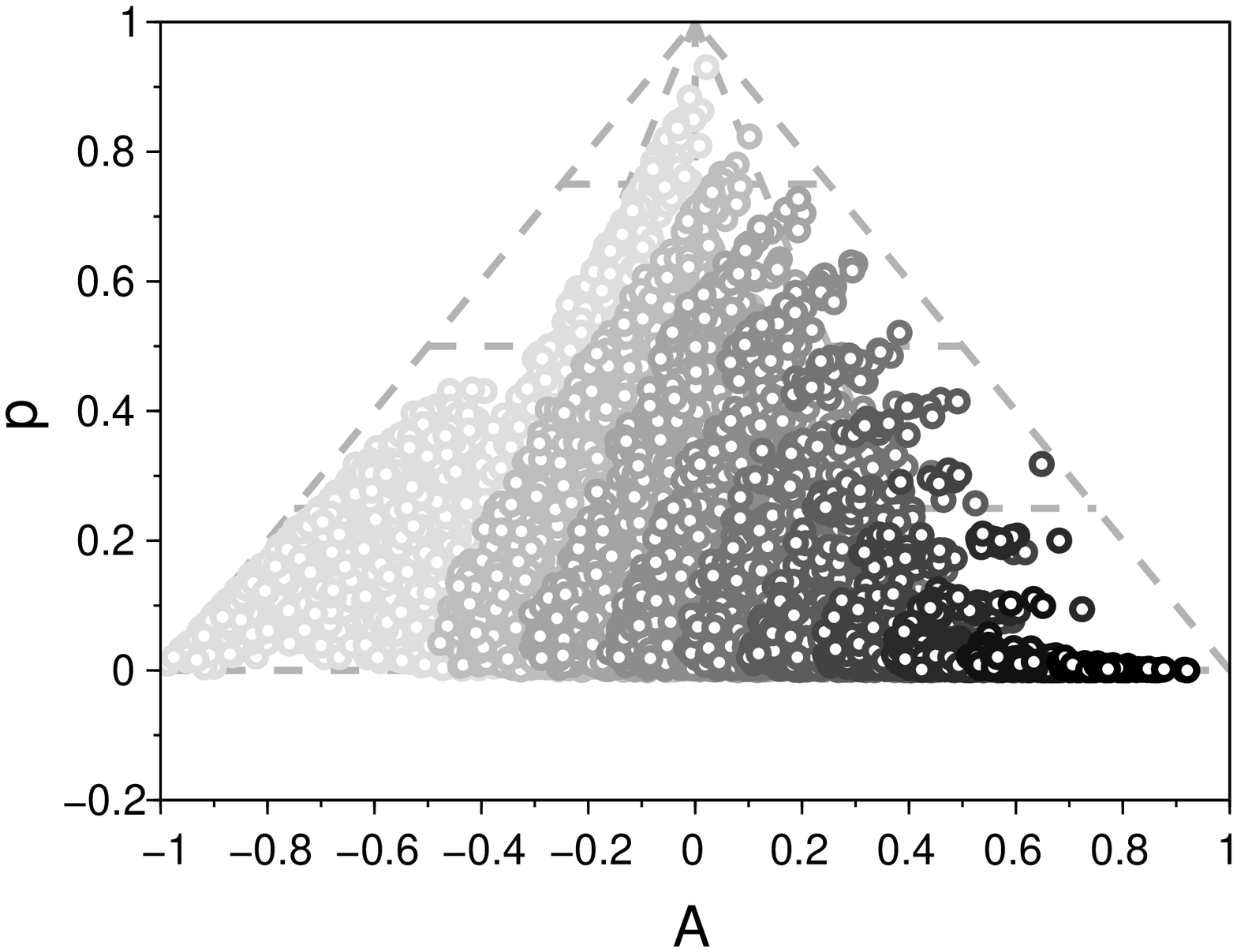}}
\caption[]{Isotropic Index for $4$-qubits random states. Probability variation.}
\label{Fig_IndIso_PROB_VarProbs_EstRand} 
\end{figure}

\begin{Remark}
Even though there is no direct relationship between the probabilities and the \emph{Isotropic Alignment}, from (\ref{eqn_IsoErr_DefIndIsoAlin}) it can be seen that there is a correlation between the increase (decrease) of the Alignment and the increase (decrease) of such probabilities. This can be observed in figure \ref{Fig_IndIso_PROB_VarProbs_EstRand}.
\end{Remark}

\section{Horodecki's isotropic state}
\label{sec_EstIsoHor}

An isotropic state \cite{Horodecki_1999} is a bipartite $2n$-qubits state, invariant under unitary transformations of the form $U \otimes U^*$, being $U$ any unitary matrix and $U^*$ its conjugate.

Considering $d = 2^n$, these states are described by a unique parameter $\alpha$ and are of the form
\begin{equation}
\label{eqn_EstIsoHor_EstIso}
\rho = \left( 1 - \alpha \right) \frac{I}{d^2} + \alpha \ketbra{\phi}{\phi}, \quad -\frac{1}{d^2-1} \leq \alpha \leq 1
\end{equation}
being
\begin{equation}
\label{eqn_EstIsoHor_EstMaxEnt}
 \ket{\phi} = \frac{1}{\sqrt{d}} \sum_{j=0}^{2^n-1} \ket{j} \otimes \ket{j}
\end{equation}
with $\ket{j}$ the canonical basis states of $d$-dimension space.

Unlike an isotropic error state, an isotropic state has, by definition, an even number of qubits. We now calculate the Isotropic Index for isotropic states with reference state $\ket{\phi}$.

If the parameter $\alpha$ is in $[0,1]$, and taking into consideration that $\ket{\phi}$ is a pure state, it is easy to note that the state is already expressed as in decomposition (\ref{eqn_IsoErr_DensDecomp}). Hence, for this case, the index is $\IsoInd{\varphi}{\rho} = \left( 1, 1 - \alpha \right)$, i.e. an \emph{Isotropic Alignment} equal to $1$ and an \emph{Isotropic Weight} equal to $1 - \alpha$. In this particular case of $\alpha$, the state can be seen as applying the depolarizing channel to the original state $\ket{\phi}$.

If, on the other hand, $\alpha$ is in the interval $[-1/(d^2-1),0)$, considering $p = 1 + \left(d^2 - 1 \right) \alpha$, the state can be expressed as
\begin{equation}
\label{eqn_EstIsoHor_AliHorodelNeg}
\rho = \left( 1 - \alpha \right) \frac{I}{d^2} + \alpha \ketbra{\phi}{\phi} = 
p \, \frac{I}{d^2} + \left( 1 - p \right) \frac{I - \ketbra{\phi}{\phi}}{d^2-1} = 
p \, \frac{I}{d^2} + \left( 1 - p \right) \rho_{N \phi},
\end{equation}
where $\rho_{N \phi}$ is the orthogonal isotropic mixed state of state $\ketbra{\phi}{\phi}$. It can be seen that, with the earlier definition, $p$, the \emph{Isotropic Weight}, is in $[0,1)$ and the \emph{Isotropic Alignment} is always $-1$.

\begin{figure}[!htb]
\centering
\includegraphics[width=0.6\textwidth,keepaspectratio=true]{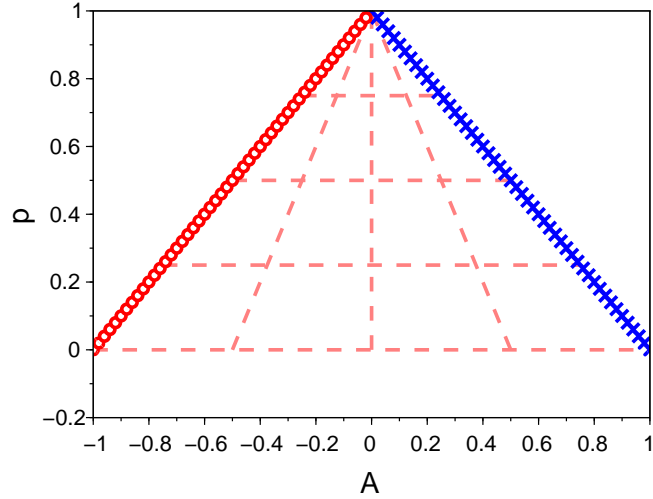}
\caption{Isotropic Index evolution for isotropic states for: $\alpha \in [0,1]$ (blue crosses) and $\alpha \in [-1/(d^2-1),0)$ (red circles).}
\label{fig_EstIsoHor_HorodeckiStates}
\end{figure}

Hence, an isotropic state can always be interpreted as an isotropic error state with reference state $\ket{\phi}$. Figure \ref{fig_EstIsoHor_HorodeckiStates} shows the evolution of the index in the triangular graph, for $\alpha \in [-1/(d^2-1),1]$.

\section{Dis-alignment in a Grover's faulty search}
\label{sec_Grover}

Grover's quantum search algorithm \cite{Grover_1996,Grover_2001} deals with the problem of finding a target element in an unstructured database of $N$ elements. It provides a quadratic speedup ($k_{Gr}=\left\lfloor \frac{\pi}{4}\sqrt{N}\right\rfloor$ steps) over the classical brute-force search.

However, considering that any quantum system is subject to error presence, the performance is affected. This phenomena has been studied by several authors in the last years \cite{Azuma_2002,Chen_2003,Shapira_2003,Regev_2008,Gawron_2012,Ambainis_2013_Grover,Temme_2014}.

In this section, the results obtained in \cite{Cohn_2016} are analyzed by using the Isotropic Index.

\subsection{Grover's algorithm}
\label{subsec_Grover_Alg}

Let there be a set of $N=2^{n}$ quantum basis states in a Hilbert space ($\mathcal{H}=\mathbb{C}^{N}$) in which the search is to be performed, and an unknown marked state (target or solution) among them. Considering a system that identifies the target (oracle), the aim is to find such target using this system in as few steps as possible.

Let $\ket{t}$ be the target element and 
\begin{equation}
\label{eq_PropagaGrover_EstIni}
	\ket{s} = \frac{1}{\sqrt{N}} \sum_{i=0}^{N-1} \ket{i}
\end{equation}
the initial state. In density matrix representation, these states are $\rho_{t}=\ketbra{t}{t}$ and $\rho_{s}=\ketbra{s}{s}$, respectively. {\bf Algorithm \ref{alg_PropagaGrover_grover}} describes this search.

\begin{algorithm}[h]
\BlankLine
\label{alg_PropagaGrover_grover}
\caption{Grover's search algorithm}
{\bf 1.} Set up the superposition state $\rho_{s}$.\\
{\bf 2.} Apply the oracle operator $O=2\left| t \right\rangle \left\langle t\right|-I$.\\
{\bf 3.} Apply the diffusion operator $D=2\left|s\right\rangle \left\langle s\right|-I$.\\
{\bf 4.} Repeat steps 2 and 3 $\left\lfloor \frac{\pi}{4}\sqrt{N}\right\rfloor -1 $
times.\\
{\bf 5.} Perform measurements in the canonical basis in each qubit. The target state will emerge with high probability as $N\gg1$.\\
\BlankLine
\end{algorithm}

Applying the algorithm $k$ times results in 
\begin{equation} 
\label{eq_rho_k_sk}
	\rho(k)  =  \ketbra{s_{k}}{s_{k}},
\end{equation}
with
\begin{eqnarray} \label{eq_rho_k}
	\ket{s_{k}} &=&  \sin\left( (2 k+1)\theta\right)\ket{t} +\cos((2 k+1)\theta)\ket{\bar{t}}, \nonumber \\ 
	\ket{\bar{t}} &=&  \frac{1}{\sqrt{N-1}}  \sum_{\substack{i=0 \\ i\neq t}}^{N-1} \ket{i} \ \, \text{and} \ \,
	\theta  =  \arcsin\left(\frac{1}{\sqrt{N}}\right).
\end{eqnarray}

Therefore, the success probability (of obtaining the target state) after $k$ steps is 
\begin{equation}\label{eq_p_k}
	p(k) = \sin^{2}\left(\left(2k+1\right)\theta\right).
\end{equation}

\begin{figure}[!htb]
\centering
\includegraphics[width=0.3\textwidth,keepaspectratio=true]{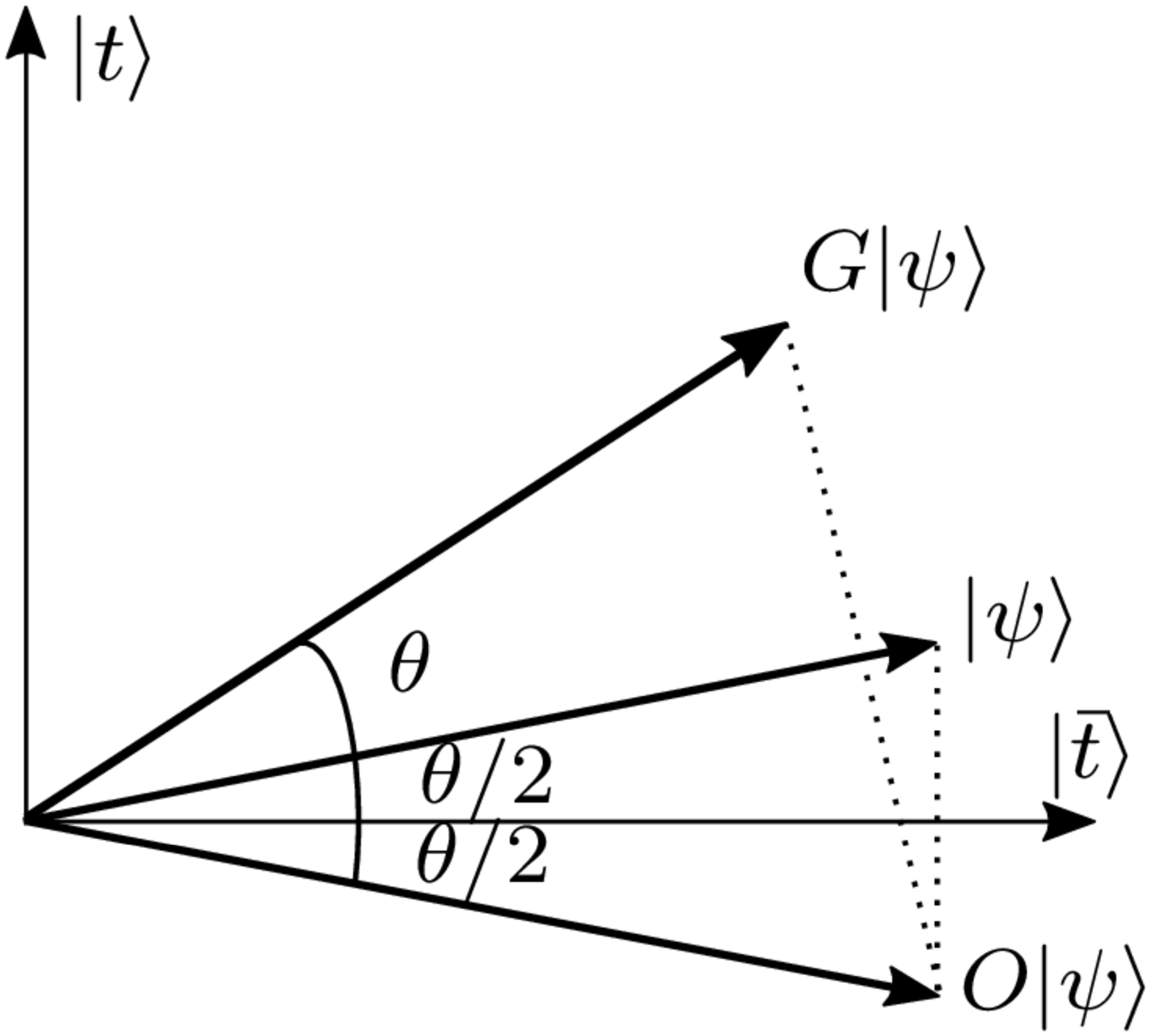}
\caption[Grover's algorithm hyperplane]{Grover's algorithm hyperplane generated by $\ket{t}$ and $\ket{\overline{t}}$.}
\label{fig_Propaga_Grover_Plano}
\end{figure}

Grover's operator $G = DO$ can be interpreted as a double reflexion, in the hyperplane generated by the orthogonal states $\ket{t}$ and $\ket{\bar{t}}$, as shown in Figure \ref{fig_Propaga_Grover_Plano}. After $k_{Gr}$ steps, the state is close to $\ket{t}$ ($\theta/2 \approx \pi/2$).

\subsection{Local and Total Depolarizing Channels}
\label{subsec_Grover_XDCh}

\begin{figure}[ht!]
	\centering
	\subfigure[Total Depolarizing Channel error model, TDCh.]{
	\label{fig_Grover_ModTDCh} 
	\includegraphics[width=0.45\columnwidth,keepaspectratio=true]{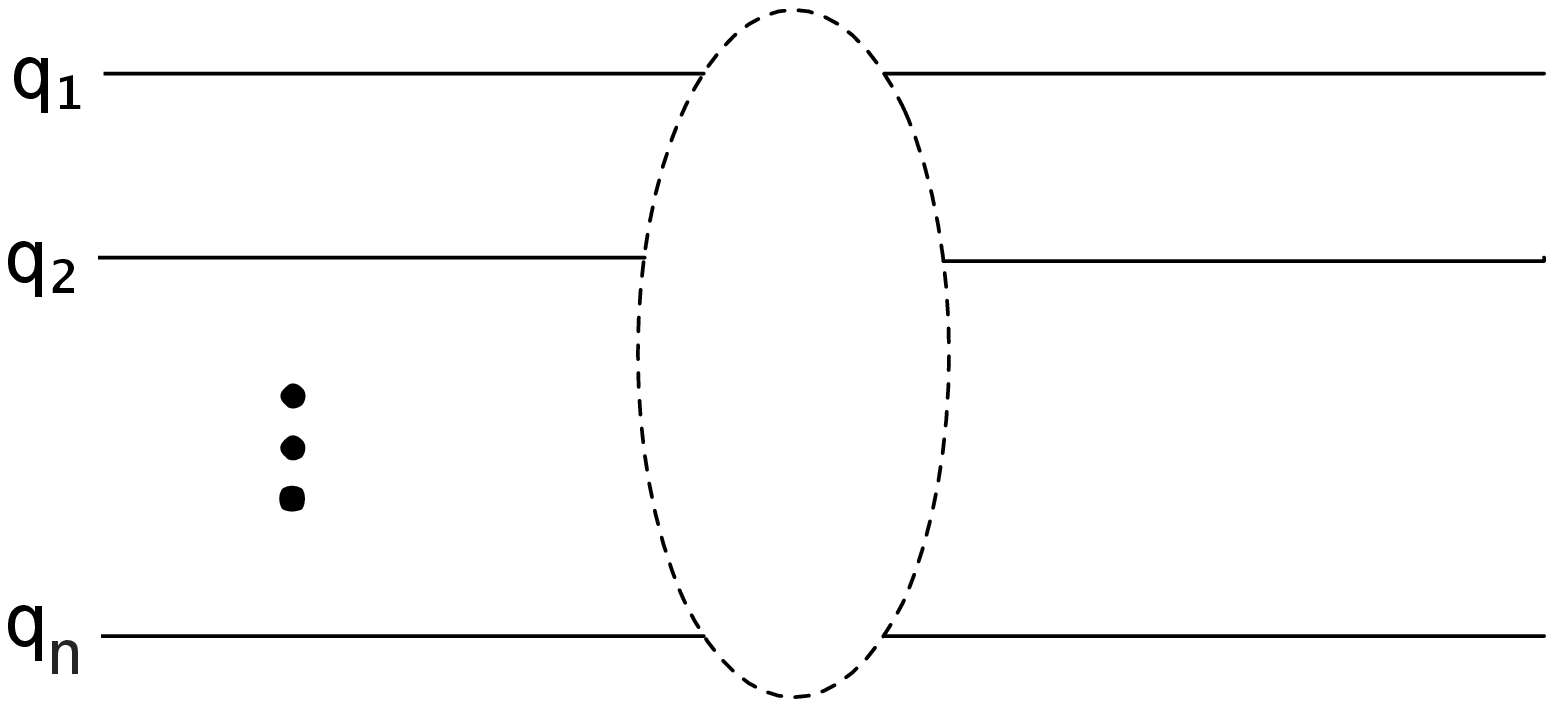}}
	\hfill
	\subfigure[Local Depolarizing Channel error model in each qubit, LDCh.]{
	\label{fig_Grover_ModLDCh} 
	\includegraphics[width=0.45\columnwidth,keepaspectratio=true]
	{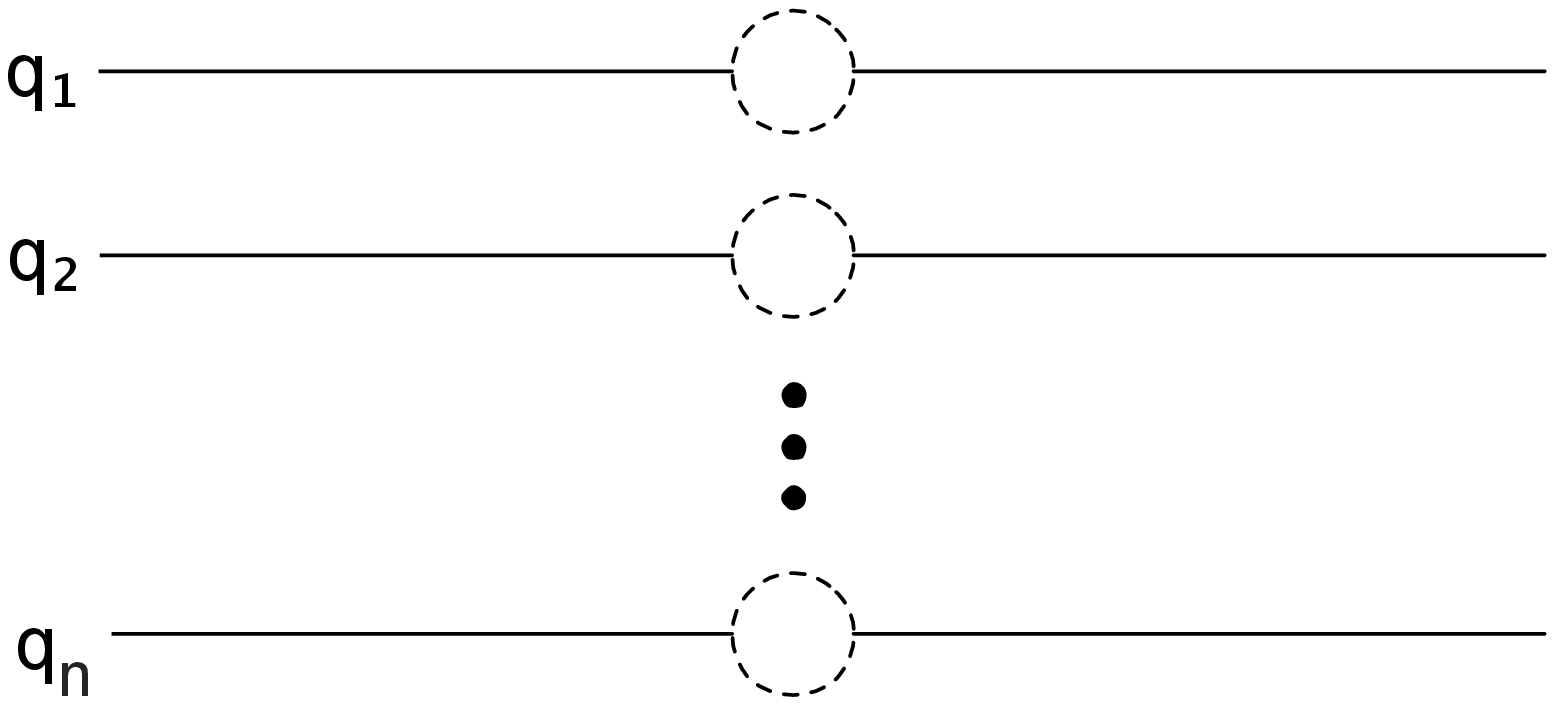}}
	\caption{Depolarizing error models. Errors are applied after each iteration of the algorithm.}
	\label{fig_Grover_ModXDCh} 
\end{figure}

Salas \cite{Salas_2008} and Vrana et al. \cite{Vrana_2014} have analyzed the degradation of the algorithm caused by the Total Depolarizing Channel (TDCh). A summary of the results obtained by Cohn et al. \cite{Cohn_2016} are presented, in which the degradation of the algorithm with TDCh and Local Depolarizing Channel (LDCh) is compared. Both error models are illustrated in Figure \ref{fig_Grover_ModXDCh}.  

The TDCh error $\varepsilon (\rho, \gamma)$, acting on a state $\rho$ with probability $\gamma$, is modeled by Eq. (\ref{eqn_IsoIndex_TriangProps3_A1}). Similarly, the LDCh error is modeled as
\begin{equation} 
\label{eq_Grover_errorLDCh}
	\varepsilon^L(\rho, \alpha)=\varepsilon_1(\rho,\alpha)\circ\varepsilon_2(\rho,\alpha) \circ \dots \circ \varepsilon_n(\rho,\alpha),	
\end{equation}
being $\alpha$ the error probability per qubit and $\varepsilon_i(\rho,\alpha)$ the depolarizing error applied to qubit $i$.

The necessary conditions to maintain the order of the original error-less algorithm are presented in Table \ref{tab_ORdenTDChLDCh} for both TDCh ($\gamma$) and LDCh ($\alpha$). As can be seen from the table, LDCh has a more restrictive condition.

\begin{table}[!h]
\centering
\begin{tabular}{cc}
\hline
 TDCh & LDCh \\ \hline
& \\
 \hspace{1cm} $\gamma \ll (k_{Gr})^{-1}$ \hspace{1cm} & \hspace{1cm} $\alpha \ll (k_{Gr} \log_2N)^{-1}$ \hspace{1cm} \\
& \\  \hline
\end{tabular}%
\caption{Necessary conditions in $\gamma$ (TDCh) and $\alpha$ (LDCh) to maintain Grover's algorithm original order $\Theta(\sqrt{N})$.} 
\label{tab_ORdenTDChLDCh}
\end{table}

\subsection{Isotropic Index interpretation}
\label{subsec_Grover_AnaIso}

The Isotropic Index provides a new insight in the analysis of the noisy algorithms evolution. In this case, the error-less state in each $k$ step (\ref{eq_rho_k}) is used as the reference state. 

\subsubsection{Total Depolarizing Channel}
\label{subsec_degradacion_subsec_TDCh}

As shown in \cite{Cohn_2016}\cite{Vrana_2014}, after $k$ steps the state is
\begin{equation}
\label{eq_grover_TDC_rho_k}
	\rho (k, \gamma) = \left[1-\left(1-\gamma\right)^k \right]\frac{I}{2^n} + \left(1-\gamma\right)^k\rho(k), 
\end{equation}
where $\rho(k)$ is the error-less density matrix (\ref{eq_rho_k_sk}) and $\gamma$ is the error probability. 

\begin{figure}[ht!]
	\centering
	\subfigure[\emph{Isotropic Weight} evolution versus steps.]{
	\label{fig_Grover_TDCh_IsoP} 
	\includegraphics[width=0.45\columnwidth,keepaspectratio=true]
	{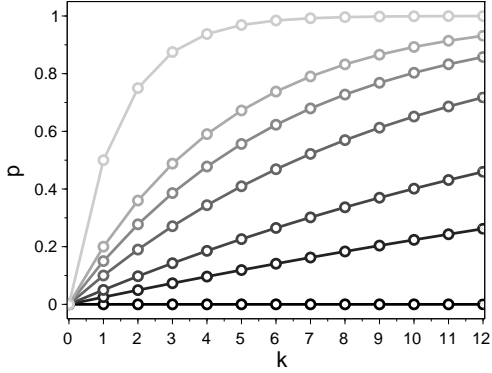}}
	\hfill
	\subfigure[Evolution in the Isotropic Triangle.]{
	\label{fig_Grover_TDCh_IsoTriang} 
	\includegraphics[width=0.45\columnwidth,keepaspectratio=true]
	{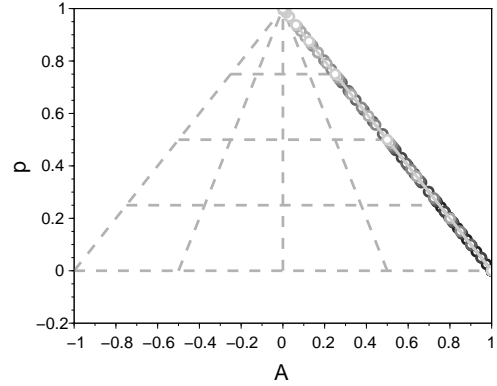}}
	\caption{Grover's algorithm with TDCh. Evolutions considering several $\gamma$ values in $\left\{0, 0.025, 0.05, 0.1, 0.15, 0.2, 0.5 \right\}$. Darker tones indicate lower error values.}
	\label{fig_Grover_TDCh_Iso} 
\end{figure}

Comparing equations (\ref{eq_grover_TDC_rho_k}) and (\ref{eqn_IsoErr_DensDecomp}), it is easy to see that the \emph{Isotropic Weight} is $p = 1 - \left(1-\gamma\right)^k$, and the \emph{Isotropic Alignment} is always $1$. Figure \ref{fig_Grover_TDCh_Iso} shows the index evolution for several $\gamma$ values. 

\subsubsection{Local Depolarizing Channel}
\label{subsubsec_degradacion_subsec_LDCh}

\begin{figure}[!htb]
\centering
\subfigure[\emph{Isotropic Alignment} evolution versus steps.]{
\label{fig_ResGrover_Ind_Iso_al} 
\includegraphics[width=0.45\columnwidth,keepaspectratio=true]{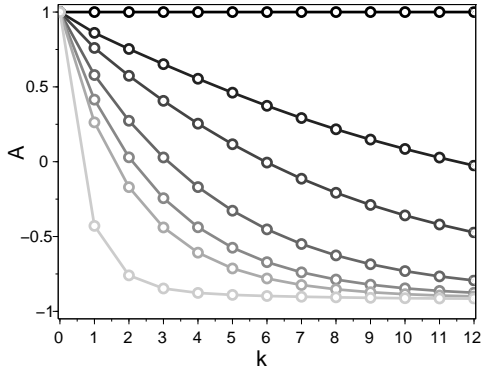}}
\hfill
\subfigure[\emph{Isotropic Weight} evolution versus steps.]{
\label{fig_ResGrover_Ind_Iso_p} 
\includegraphics[width=0.45\columnwidth,keepaspectratio=true]{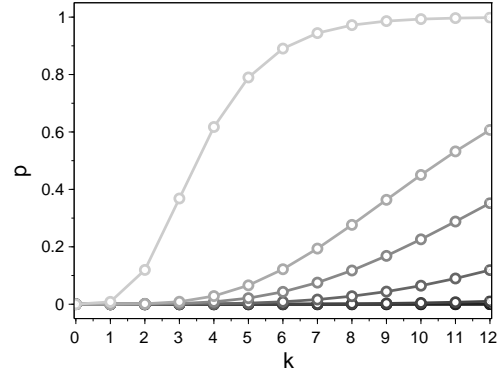}}
\subfigure[Evolution in the Isotropic Triangle.]{
\label{fig_ResGrover_Ind_Iso_ind} 
\includegraphics[width=0.45\columnwidth,keepaspectratio=true]{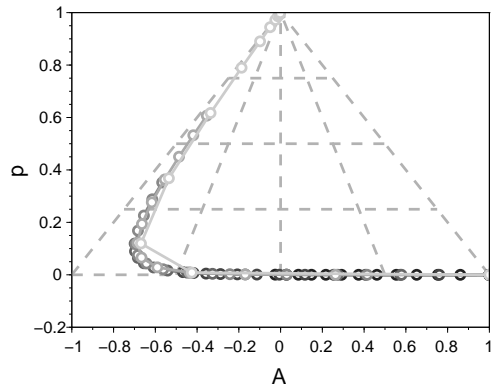}}
\caption{Grover's algorithm with LDCh. Evolutions considering several $\alpha$ values in $\left\{0, 0.025, 0.05, 0.1, 0.15, 0.2, 0.5 \right\}$. Darker tones indicate lower error values.}
\label{fig_ResGrover_Ind_Iso} 
\end{figure}

As seen above, Grover's algorithm is more sensitive to the LDCh error rather than to the TDCh. In the former, even for small values of \emph{Isotropic Weight} $p$, the \emph{Isotropic Alignment} has a huge deviation towards $-1$ (i. e. close to $\rho_{N \varphi}$), as illustrated in figure \ref{fig_ResGrover_Ind_Iso}. This fact can be interpreted as an indicator of the state deviation from the original hyperplane (figure \ref{fig_Propaga_Grover_Plano}), which affects directly the order.

\section{Error propagation in Shor's code}
\label{sec_Shor}

In the last decades, with the purpose of protecting information from errors inherently present in any quantum system, diverse proposals have appeared \cite{Calderbank_1996,Knill_1998,Knill_2000}. Error-correcting codes \cite{Laflamme_1996} enabled the development of, among others, fault-tolerant designs \cite{Gottesman_1998,Aharonov_2008,Gottesman_2010}. In this section, we analyze the performance of Shor's $9$-qubit code \cite{Shor_1995} under the effect of the LDCh error using the Isotropic Index.

\subsection{Shor's code}
\label{subsec_Shor_Codigo}

\begin{figure}[!ht]
\centering
\includegraphics[width=0.7\columnwidth,keepaspectratio=true]{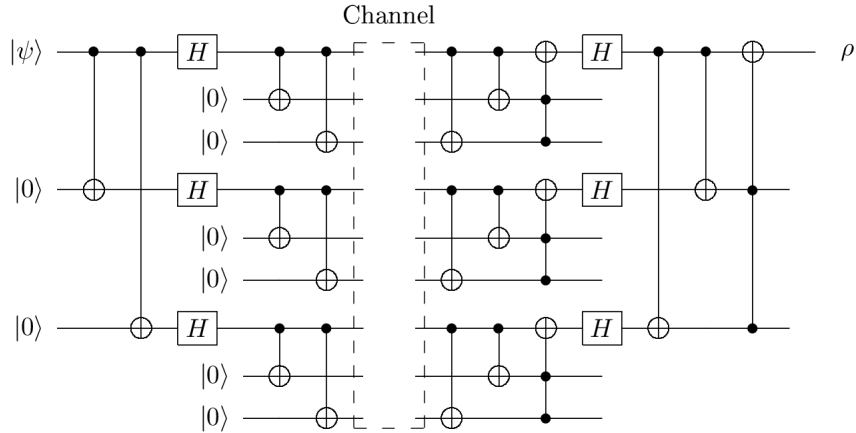}
\caption{Shor's code. Encoder/Channel/Decoder\&Corrector.}
\label{fig_CT2016_PECC_CodShor_ShorCod}
\end{figure}

The $9$-qubit code, proposed by Peter Shor in 1995, is one of the first proposals for quantum error correction. It is capable of correcting any kind of error in any qubit, as long as there is an error present in only one qubit. It does so by combining $3$-qubits codes for both logical and phase errors. The circuit, as illustrated in figure \ref{fig_CT2016_PECC_CodShor_ShorCod}, shows a particular unitary implementation that does not need to measure syndrome ancillas.

The corresponding logical states are 
\begin{eqnarray}
   \ket{\psi} = \ket{0} \rightarrow \ket{0_L} & \equiv & \frac{1}{\sqrt{8}} \left( \ket{000} + \ket{111} \right) ^ {\otimes 3}, \nonumber \\
   \ket{\psi} = \ket{1} \rightarrow \ket{1_L} & \equiv & \frac{1}{\sqrt{8}} \left( \ket{000} - \ket{111} \right) ^ {\otimes 3}.
\label{eqn_CT2016_PECC_CodShor_ShorWords}
\end{eqnarray}

\subsection{Error propagation analysis}
\label{subsec_Shor_Errores}

\begin{figure}[!htb]
	\centering
	\subfigure[$P_{ex}$ for every initial state $\ket{\psi}$ in Bloch's sphere. Error probability $\alpha = 0.1$.]{
	\label{fig_Shor_LDCh_CanalEsfera} 
	\includegraphics[width=0.45\textwidth,keepaspectratio=true]
	{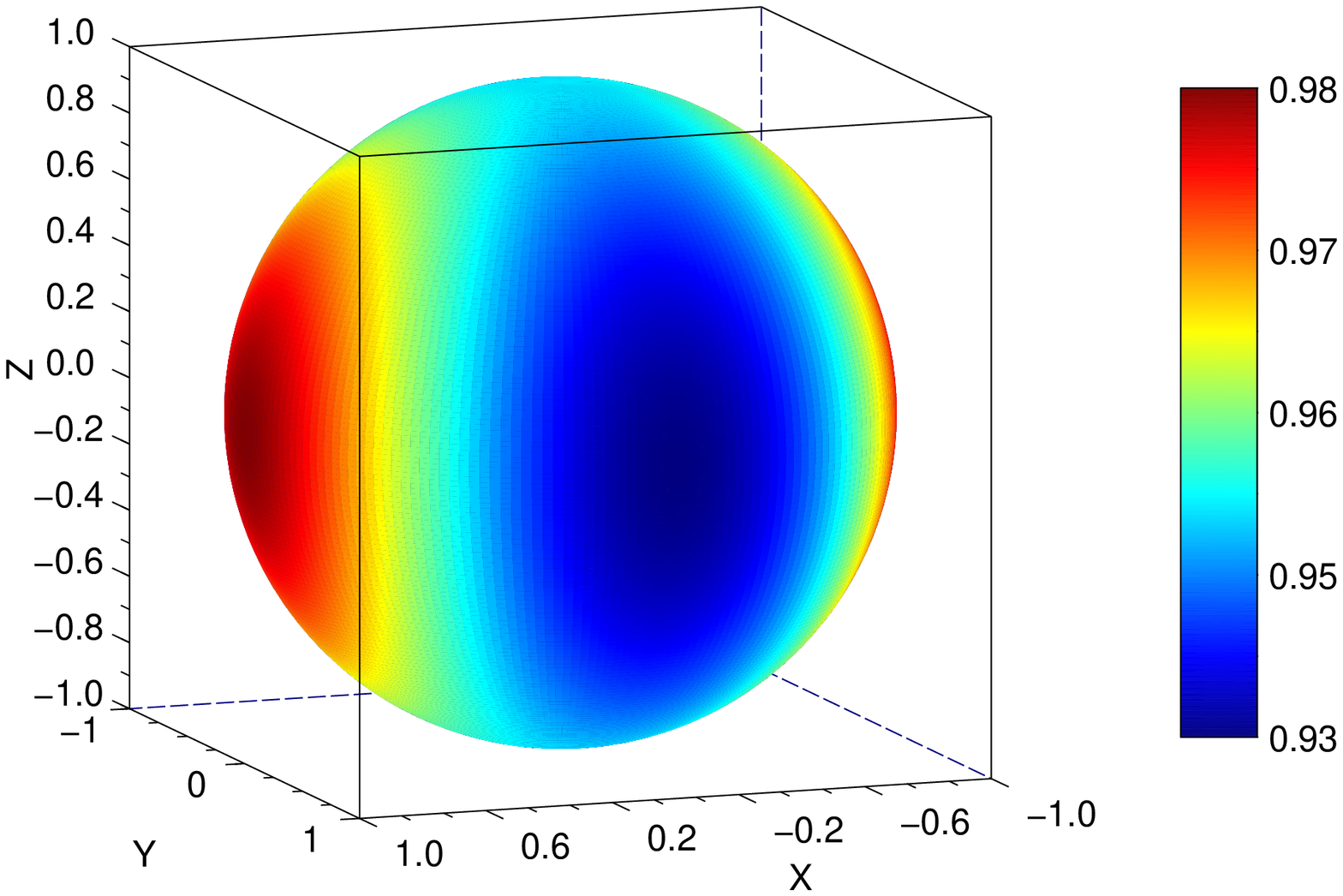}} %
	\hfill
	\subfigure[$P_{ex}$ as a function of channel error probability $\alpha$. Solid black curve corresponds to an initial state $\ket{+}$, dashed dark gray to $\ket{0}$ and dotted light gray to $\ket{y_+}$.]{
	\label{fig_Shor_LDCh_CanalCurvas} 
	\includegraphics[width=0.45\textwidth,keepaspectratio=true]
	{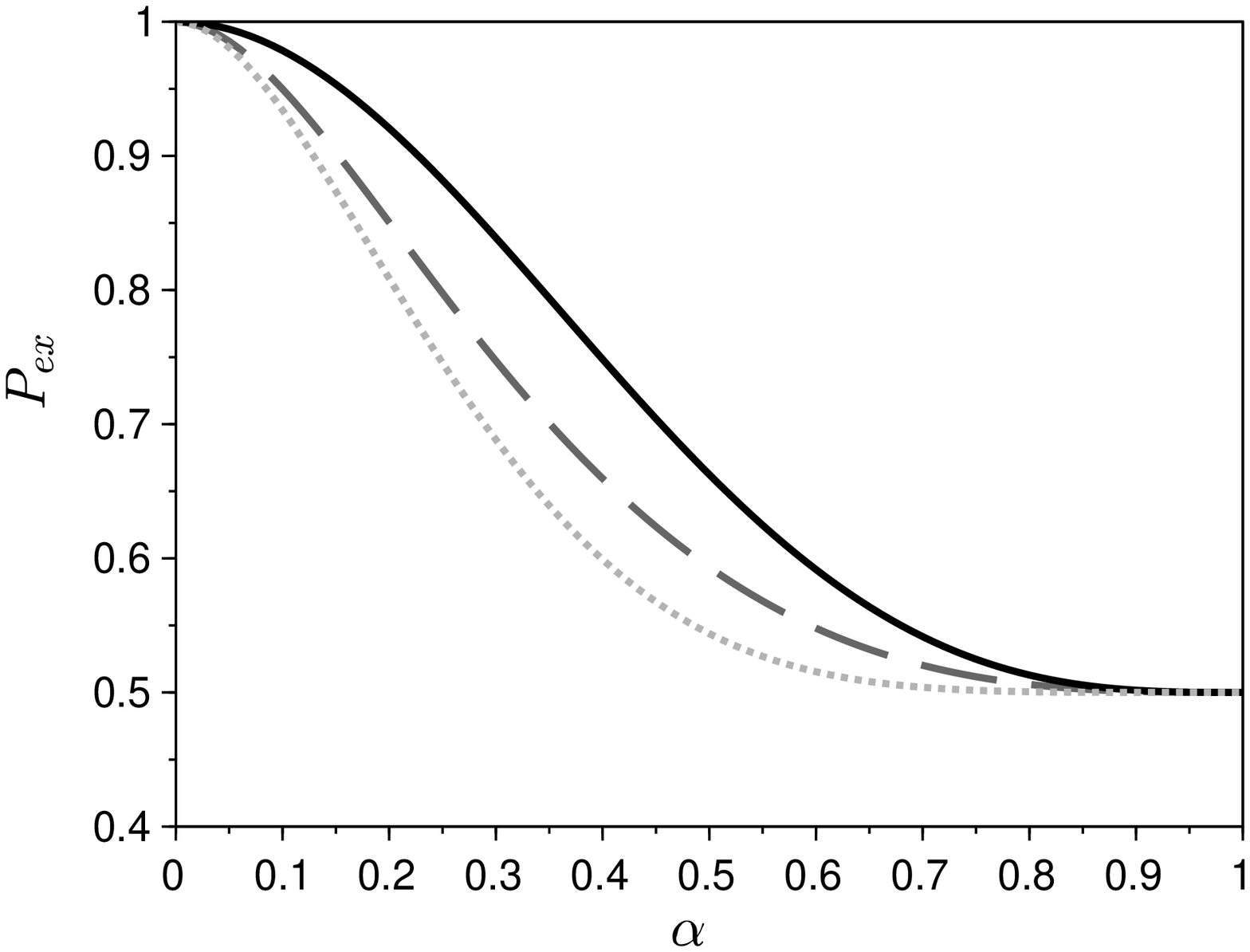}}
	\caption{Success probability $P_{ex}$ for different initial states.}
	\label{fig_Shor_LDCh_Canal} 
\end{figure}

In this analysis, we will consider perfect quantum gates and model the error in the channel as a LDCh in each qubit, with $\alpha \in [0,1]$ being the error probability, as in (\ref{eq_Grover_errorLDCh}).
We define the success probability (probability of obtaining the initial state $\ket{\psi}$) as 
\begin{equation}
\label{eq_CT2016_PECC_AnalisaShor_Canal_ProbEx}
P_{ex} = Fid^2 \left( \ket{\psi}, \rho \right) = \bradensket{\psi}{\rho}{\psi}.
\end{equation}
It can be seen, figure \ref{fig_Shor_LDCh_CanalEsfera}, that the states with higher probability are 
\begin{equation}
\label{eq_CT2016_PECC_AnalisaShor_Canal_MasProbables}
\ket{+} = \frac{ \ket{0} + \ket{1} }{\sqrt{2}} \quad \textrm{and} \quad \ket{-} = \frac{\ket{0} - \ket{1}}{\sqrt{2}},
\end{equation}
whereas the states with lower probability are
\begin{equation}
\label{eq_CT2016_PECC_AnalisaShor_Canal_MenosProbables}
\ket{y_+} = \frac{\ket{0} + i \ket{1}}{\sqrt{2}} \quad \textrm{and} \quad \ket{y_-} = \frac{\ket{0} - i \ket{1}}{\sqrt{2}}.
\end{equation}
This result is valid for all $\alpha \in (0,1)$.

Figure \ref{fig_Shor_LDCh_CanalCurvas} illustrates the variation of the success probability $P_{ex}$ as a function of error probability $\alpha$ for the initial states $\ket{+}$, $\ket{0}$ and $\ket{y_+}$. 

In order to analyze the results using the Isotropic Index, it is necessary to decompose the final state $\rho$ as in equation (\ref{eqn_IsoErr_DensDecomp}). In this case, success probability $P_{ex}$ is given by
\begin{equation}
\label{eq_CT2016_PECC_AnalisaShor_CanalGamma_ProbInd}
P_{ex} = \frac{p}{2} + (1-p) \bradensket{\psi}{\hat{\rho}}{\psi},
\end{equation}
with $p$ being the \emph{Isotropic Weight}. 

\begin{figure}[!htb]
\centering
\includegraphics[width=0.6\columnwidth,keepaspectratio=true]{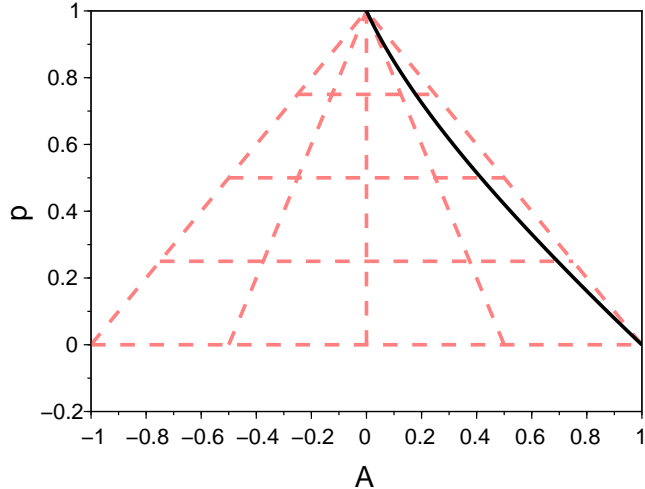}
\caption{Isotropic Index evolution for $\alpha \in [0, 1]$ and initial state $\ket{\psi_{\pi/4}}$.}
\label{fig_CT2016_PECC_AnalisaShor_CanalGamma_DeltayGamma_Triang}
\end{figure}

Figure \ref{fig_CT2016_PECC_AnalisaShor_CanalGamma_DeltayGamma_Triang} shows the Isotropic Index with initial state 
\begin{equation}
\ket{\psi_{\pi/4}} = (1/\sqrt{2}) \left(\ket{0} + e^{i \frac{\pi}{4}} \ket{1} \right),
\label{eq_psi_pi_4}
\end{equation}
which is the state that has the lowest value of \emph{Isotropic Alignment}. As can be observed in the figure this value is close to $1$. Hence, for any initial state $\ket{\psi}$, the \emph{Isotropic Alignment} is always close to $1$. Therefore, by equation (\ref{eqn_IsoErr_DefIndIsoAlin}), the term $\bradensket{\psi}{\hat{\rho}}{\psi}$ is close to $A^2$, and a zero-order approximation yields
\begin{equation}
\label{eq_CT2016_PECC_AnalisaShor_CanalGamma_ProbInd_Approx}
P_{ex} \approx 1-\frac{p}{2}.
\end{equation}
Finally, it is relevant to note that the loss of probability is mainly given by the \emph{Isotropic Weight} and not by the dis-alignment.

\section{Conclusion}
\label{sec_Conclusiones}

In this paper we introduce an (double) Isotropic Index $Iso(A,p)$ in order to study error propagation in quantum algorithms, given an initial pure reference state. This index characterizes the isotropy using two components: an inherent isotropic component, or \emph{Isotropic Weight}, $p$ that indicates the maximal mixed $I/2^n$ part; and a residual component, that quantified by the \emph{Isotropic Alignment} indicates the deviation from an actual isotropic error state. 

Some examples have been shown in order to explain that degradation due to error propagation may have different origins: some caused by the accumulation of the isotropic component (Shor's code) and others by the increase of deviation (Grover's algorithm).

\section{Acknowledgment}
\label{sec_Gracias}

A.L.F.O. and E.B. acknowledge financial support from SNI-Uruguay.

\bibliographystyle{unsrt}      

\bibliography{QIP_2016}

\end{document}